\documentclass[10pt,aps,prl,twocolumn,superscriptaddress]{revtex4-1}
\usepackage{graphicx}
\usepackage{color}
\usepackage{amsfonts}
\usepackage{amsmath}
\usepackage[percent]{overpic}
\let\oldAA\AA
\renewcommand{\AA}{\text{\normalfont\oldAA}}

\DeclareRobustCommand{\rchi}{{\mathpalette\irchi\relax}}
\newcommand{\irchi}[2]{\raisebox{\depth}{$#1\chi$}} 
\usepackage{changes}

\begin{document}
\title{Transient Hot Electron Dynamics in Single-Layer TaS$_2$}
\author{Federico Andreatta}
\altaffiliation{Contributed equally}
\affiliation{Department of Physics and Astronomy, Interdisciplinary Nanoscience Center, Aarhus University,
8000 Aarhus C, Denmark}
\author{Habib Rostami}
\altaffiliation{Contributed equally}
\affiliation{Nordita, 106 91 Stockholm, Sweden}
\author{Antonija Grubi\v{s}i\'{c} \v{C}abo}
\author{Marco Bianchi}
\affiliation{Department of Physics and Astronomy, Interdisciplinary Nanoscience Center, Aarhus University,
8000 Aarhus C, Denmark}
\author{Charlotte E. Sanders}
\affiliation{Central Laser Facility, STFC Rutherford Appleton Laboratory, Harwell, United Kingdom}
\author{Deepnarayan Biswas}
\affiliation{SUPA, School of Physics and Astronomy, University of St. Andrews,
St. Andrews, United Kingdom}
\author{Cephise Cacho}
\author{Alfred J. H. Jones}
\author{Richard T. Chapman}
\author{Emma Springate}
\affiliation{Central Laser Facility, STFC Rutherford Appleton Laboratory, Harwell, United Kingdom}
\author{Phil D. C. King}
\affiliation{SUPA, School of Physics and Astronomy, University of St. Andrews,
St. Andrews, United Kingdom}
\author{Jill A. Miwa}
\affiliation{Department of Physics and Astronomy, Interdisciplinary Nanoscience Center, Aarhus University,
8000 Aarhus C, Denmark}
\author{Alexander Balatsky}
\affiliation{Nordita, 106 91 Stockholm, Sweden}
\affiliation{Institute for Materials Science, Los Alamos National Laboratory, Los Alamos, New Mexico 87545, USA}
\author{S{\o}ren~Ulstrup}
\affiliation{Department of Physics and Astronomy, Interdisciplinary Nanoscience Center, Aarhus University,
8000 Aarhus C, Denmark}
\author{Philip~Hofmann}
\email{philip@phys.au.dk}
\affiliation{Department of Physics and Astronomy, Interdisciplinary Nanoscience Center, Aarhus University,
8000 Aarhus C, Denmark}

\date{\today}
\begin{abstract}
Using time- and angle-resolved photoemission spectroscopy, we study the response of metallic single layer TaS$_2$ in the 1H structural modification to the generation of excited carriers by a femtosecond laser pulse. A complex interplay of band structure modifications and electronic temperature increase is observed and analyzed by direct fits of model spectral functions to the two-dimensional (energy and $k$-dependent) photoemission data. Upon excitation, the partially occupied valence band is found to shift to higher binding energies by up to 150 meV, accompanied by electronic temperatures exceeding 3000~K. These observations are explained by a combination of temperature-induced shifts of the chemical potential, as well as temperature-induced changes in static screening. Both contributions are evaluated in a semi-empirical tight-binding model. The shift resulting from a change in the chemical potential is found to be dominant.
\end{abstract}

\maketitle



\section{Introduction}
Two-dimensional (2D) electron systems can be dramatically altered and driven into a number of distinct phases by the application of external fields. A prime example of this is the ability to change the electron-electron interaction by electric field control of the charge carrier density in 2D electron gasses (2DEGs) confined in high electron mobility transistors \cite{Smith:1972,Shashkin:2003}. Tuning such systems into extreme conditions can create an electronic instability such as a metal-insulator transition \cite{Pudalov:2002} or lead to the emergence of superconductivity \cite{Wan:2015}. Even when merely doped by using different substrates or adsorbates, Fermi level shifts on the order of electron volts can be induced in semimetallic 2D materials such as graphene \cite{Bostwick:2007aa,Pletikosic:2009aa,Larciprete:2012aa,Ulstrup:2014ad}, providing access to the dependence of the electronic self-energy on the electron/hole density over a wide energy range \cite{Ulstrup:2016ab}. In principle, the electronic self-energy also depends on the temperature but such effects are usually negligible because $k_B T$ at reachable temperatures tends to be much smaller than the electronic bandwidth \cite{Giuliani:2005,Zhang:2004ab}. The emergence of interesting physics therefore requires narrow electronic bandwidths or high electronic temperatures. These conditions are ideally reached in 2D systems where non-trivial temperature effects are expected to influence density-dependent many-body effects \cite{Coleridge:1996,Shin:1998}. 

By employing an intense optical excitation in a pump-probe scheme such as time- and angle-resolved photoemission (TR-ARPES) to drive (semi) metallic 2D materials out of equilibrium, an extremely wide range of transient electronic temperatures can be accessed \cite{Johannsen:2013ab}. This approach is effective for studying instabilities in the electronic system under extreme conditions and on ultrafast time scales where transient charge order effects and metal-insulator transitions may completely alter the electronic spectrum around the Fermi energy as observed in metallic transition metal dichalcogenides (TMDCs) \cite{Perfetti:2006,Petersen:2011,Rohwer:2011,Hellmann:2012,Monney2016}. 

Here we probe the electronic response of a metallic 2D TMDC, single layer (SL) TaS$_2$ of the 1H polymorph, to the excitation of electrons by a femtosecond laser pulse with a photon energy of 2.05~eV using TR-ARPES. We observe that this leads to a strongly excited thermal distribution of the electrons within the time resolution of our experiment, with electronic temperatures exceeding 3000~K. Interestingly, the elevated electronic temperature is accompanied by a binding energy shift of the band structure of up to $\approx$100~meV.  Such a change of the electronic structure is not necessarily expected for a 2D metal. While dynamic band structure renormalization is not uncommon in 2D semiconductors due to the strongly enhanced screening by degenerate transient doping \cite{Steinhoff:2014aa,Chernikov:2015aa,Ulstrup:2016a}, and complex band shifts have even been observed for bulk insulators \cite{Mor:2017aa}, such effects are absent in semimetallic systems such as graphene \cite{Johannsen:2013ab} or bilayer graphene \cite{Ulstrup:2014ac}.
  
This paper is structured as follows: this Introduction section is followed by an Experimental section that provides the details of the sample preparation, static ARPES, and TR-ARPES. The Results and Discussion section is divided into five subsections that give (i) a presentation of the experimental results, (ii) a description of an approach to fit the full $(E,k)$-dependence of the measured photoemission intensity, (iii) a presentation of the band shifts and electronic temperatures resulting from our analysis, (iv) a theoretical description of the expected band shifts in the single particle picture, and (v) a section accounting for the effect of temperature-dependent screening. Finally, the main results and their implications are summarized in a Conclusions section. 

\section{Experimental}

 SL TaS$_2$ was grown on bilayer graphene by ultra-high vacuum (UHV) van der Waals epitaxy, in a manner similar to that used previously for the growth of SL MoS$_2$ \cite{miwavander2015}. Tantalum from an e-beam evaporator was deposited on a graphene bilayer on top of a buffer layer on the Si-face of a 6H-SiC(0001) substrate (TanKeBlue Semiconductor Co.) \cite{Yu:2011aa}. The Ta deposition took place in a H$_2$S atmosphere of $\sim$10$^{-6}$~mbar for two minutes. Subsequently, the sample was annealed for 20 minutes at  590~K in the same background pressure of H$_2$S. Repetition of this procedure allowed for an increase in coverage. For the sample used in this experiment, two growth cycles were performed, resulting in a coverage on the order of one monolayer, as judged by the reduced photoemission intensity from the graphene $\pi$-band.  Due to the weak interaction between the SL TaS$_2$ and the underlying bilayer graphene, TaS$_2$ domains are found to be randomly oriented with respect to the substrate; this is clearly evidenced by both low energy electron diffraction (LEED) and scanning tunnelling microscopy (STM). Indeed, the interaction is so weak that individual domains can be moved on the surface using the tip of the STM. 

The growth of SL TaS$_2$ on bilayer graphene and the resulting electronic structure were monitored using core level spectroscopy  \cite{suppl} and ARPES at the SGM3 beamline at the ASTRID2 synchrotron light source using a photon energy of 25~eV at a sample temperature of $\approx$44~K \cite{Hoffmann:2004aa}. The energy and angular resolution were set to 25~meV and 0.2$^{\circ}$, respectively. Since the typical size of the  SL TaS$_2$ islands (less than 10~nm)  is much smaller than the spot size of the synchrotron beam (100~$\mu$m), the random orientation of the islands results in the observation of an azimuthally averaged band structure, in contrast to the well-defined Dirac cone of the underlying bilayer graphene  \cite{suppl}. 

TR-ARPES experiments were performed at the Artemis facility, Rutherford Appleton Laboratory \cite{Frassetto:2011}. The sample was transferred from the growth chamber at ASTRID2 to the TR-ARPES facility under UHV conditions, which was necessary because the chemical sensitivity of TaS$_2$ causes degradation of the samples if they are exposed to air. A 1~kHz Ti:sapphire amplified laser system with a fundamental wavelength of 785~nm was used to generate $p$-polarized high harmonic probe pulses with an energy of 25~eV for photoemission and $s$-polarized pump pulses with a photon energy of 2.05~eV for optical excitation of our sample. The angular and time resolution were 0.3$^{\circ}$ and 40~fs, respectively, while the energy resolution varied between 300 and 800~meV depending on the beamline and detector settings and the fluence of the optical pulses. Different fluences and sample temperatures were investigated. The spot size of the laser beams are also of the order of 100 $\mu$m, leading to the same averaging of the electronic structure from azimuthally disordered SL TaS$_2$ areas as for the experiments using the spot coming from the synchrotron light source.

\section{Results and Discussion}
\subsection{(i) Time-resolved ARPES on TaS$_2$}

\begin{figure} [t!]
\begin{center}
\includegraphics{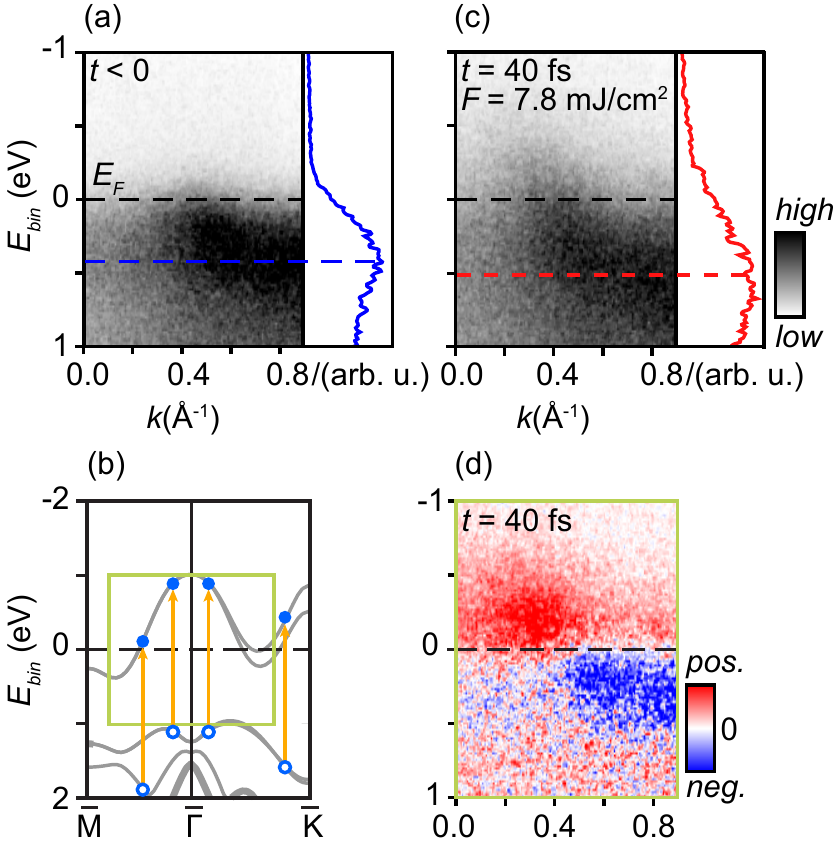}
\caption{(Color online) TR-ARPES measurement of the SL TaS$_2$ dispersion around $E_F$: (a) Left: Measured spectrum before optical excitation ($t<0$), sample temperature 300~K. Right: Energy distribution curve taken at the $k$ value of the band's highest binding energy as given in the text. The dashed red line is an estimate of the peak position. (b) Calculated dispersion from Ref. \citenum{Sanders2016} with examples of possible direct electron (filled circles) and hole (open circles) excitation processes (arrows). The region enclosed by a green square marks the $(E,k)$-space probed in the TR-ARPES experiment. (c) TR-ARPES data as in (a) but at the peak of optical excitation ($t=40$~fs). Right: Energy distribution curve taken as in (a). (d) Difference spectrum: intensity difference obtained by subtracting the intensity for $t<0$ in (a) from that at $t=40$~fs in (c).}
\label{fig:1}
\end{center}
\end{figure}

The electronic structure of undoped SL TaS$_2$ is characterized by a half-filled band with a Fermi surface consisting of hole pockets around the $\bar{\Gamma}$ and $\bar{K}$ points of the 2D Brillouin zone (BZ) \cite{Sanders2016}. A TR-ARPES spectrum taken near $\bar{\Gamma}$  without optical pumping is shown in Fig. \ref{fig:1}(a). It shows a dispersive feature with a highest binding energy of  $\approx$400~meV at $\approx$0.8~\AA$^{-1}$. It crosses the Fermi level at  $\approx$0.5~\AA$^{-1}$ and   is unoccupied at $\bar{\Gamma}$. In view of the azimuthal disorder between different domains of SL TaS$_2$ on the sample, this observed  dispersion does not correspond to any particular high-symmetry direction in the calculated band structure of Fig. \ref{fig:1}(b),  but has to be interpreted as an average over all possible orientations, roughly corresponding to an average over the $\bar{\Gamma}-\bar{M}$ and $\bar{\Gamma}-\bar{K}$ directions marked in Fig. \ref{fig:1}(b) \cite{Sanders2016}. The main differences between these directions are the higher maximum binding energy reached along $\bar{\Gamma}-\bar{M}$ and the strong spin-orbit splitting along $\bar{\Gamma}-\bar{K}$. However, for small $|k|$, the dispersion in the two directions is sufficiently similar for the average to still show the hole pocket character.  The main effect of integrating over all directions is  a broadening of the features at higher binding energies in the data shown. The spin-orbit splitting along the $\bar{\Gamma}-\bar{K}$ direction has no direct consequence for the observed band, since the upper spin-split branch is above the Fermi energy.

Figure \ref{fig:1}(c) shows the result of optically pumping the system with a photon energy of 2.05~eV at a fluence of $F=$7.8~mJ/cm$^{2}$, displaying TR-ARPES data collected at a time delay of 40~fs between the pump and the probe pulse, corresponding to the peak excitation of the system. Pumping leads to drastic changes in the  spectrum: The observed dispersion now extends well above the Fermi energy, indicating the presence of hot electrons. This is clearly seen  when considering the difference between the excited spectrum and the equilibrium spectrum in Fig. \ref{fig:1}(d). Excited carriers are expected to be generated from direct optical transitions involving occupied valence band and unoccupied conduction band states, as indicated by the arrows in Fig. \ref{fig:1}(b). The presence of a continuous distribution of hot electrons is then merely indicative of a very fast thermalization of these excited carriers that takes place at a timescale lower than our temporal resolution. 

A much more surprising result of the optical excitation is that the entire dispersion is shifted to higher binding energies by $\gtrsim$100~meV, as seen clearly by the shift between the maxima of the energy distribution curves (EDCs) in Fig. \ref{fig:1}(a) and (c). Such shifts are not necessarily expected for metallic systems. Indeed, for a simple free-electron like 2D system, one might not expect any shift at all because of the energy-independent density of states.

\subsection{(ii) Fit to simulated model spectral function}
A quantitative comparison of the observed effects to calculations requires an accurate determination of the band structure changes and the electronic temperature as a function of pump fluence and time delay. The complexity of the situation and the many unknown parameters render the conventional approach of fitting energy or momentum distribution curves by simple models impractical. Indeed, extracting the electronic temperature from such fits is already problematic even for very simple situations  \cite{Kroger:2001aa,Crepaldi:2012aa,Ulstrup:2014aa}. Instead, we introduce an approach in which the energy and $k-$dependent photoemission intensity, such as in Fig. \ref{fig:1}(a), is fitted to a model based on a resolution-broadened spectral function that can, in principle, include the single particle dispersion, many-body effects and the electronic temperature. 

The photoemission intensity measured in an ideal ARPES experiment is given by 
\begin{eqnarray}
\mathcal{I}(E,\mathbf{k})\propto |\mathcal{M}_{if}^{\mathbf{k}}|^2 \mathcal{A}(E,\mathbf{k})f(E,T), 
\label{eqn:1}
\end{eqnarray} 
where $\mathcal{A}(E,\mathbf{k})$ is the hole spectral function, $f(E,T)$ is the Fermi-Dirac distribution and $\mathcal{M}_{if}^{\mathbf{k}}$ is the energy- and $k-$dependent matrix element for the transition from the initial state $i$ to the final state $f$. If, as in the present case, the data are only collected for a small range of energy and $k$ and for a fixed photon energy and polarization, $\mathcal{M}_{if}^{\mathbf{k}}$ is expected to vary only weakly. In our experiment, the finite energy and $k-$resolution of the setup must be taken into account, such that the actual measured intensity is modelled by a convolution of $\mathcal{I}(E,\mathbf{k})$ with the appropriate resolution functions $G(\Delta E) $ and $G(\Delta k)$, assumed to be Gaussian. The measured intensity thus needs to be fitted to $\mathcal{I}_{conv}(E,k) = \mathcal{I}(E,k) \ast G(\Delta E) \ast  G(\Delta k)$. 

The azimuthally averaged photoemission intensity from SL TaS$_2$ is phenomenologically modelled as
\begin{align}
\mathcal{I}_{\mathrm{TaS_2}}(E,k) &=  \left(\mathcal{O} + \mathcal{P}E + \mathcal{Q}k\right) 
\nonumber\\ &\times \frac{\pi^{-1} \left(\alpha + \beta E + \gamma E^2 \right)  }{(E-\left(ak^2 +bk +c\right))^2 + (\alpha + \beta E + \gamma E^2 )^2} \nonumber\\ 
&\times (e^{(E-E^{\prime}_F)/k_BT_e}+1)^{-1},
\label{eqn:4}
\end{align}
where $E^{\prime}_F$ is the experimentally determined Fermi energy in the spectrum and the three parameters $\mathcal{O},\mathcal{P},\mathcal{Q}$ in the first term are used to match the calculated photoemission intensity to the experimental results, allowing for the possibility of a linear dependence on $E$ and $k$ that could arise from, \emph{e.g.}, small matrix element variations. The second term represents the spectral function in which the single particle dispersion is approximated by a parabola as  $ak^2 +bk +c$, with its minimum position constrained to $k_{min} = b/2a = 0.81$~$\AA^{-1}$, as determined from a high-resolution ARPES spectrum. The electronic self-energy  appearing in the spectral function is a complex quantity with the real part re-normalizing the dispersion and the imaginary part $\Gamma$ broadening the features. Here we assume that the real part is zero and that $\Gamma=\alpha + \beta E +\gamma E^{2}$. This will always result in an increased broadening at higher energies that accounts for the azimuthal averaging over the somewhat anisotropic band structure. Care should thus be taken, to assign physical significance to $\Gamma$.
In the third term, the population of the states is dictated by the Fermi-Dirac distribution with $T_e$ referring to the electronic temperature in the SL TaS$_2$.  In addition to the description of the  SL TaS$_2$ spectral function, it is necessary to account for the background intensity which is described in further detail in the supporting information \cite{suppl}. 

\begin{figure}
\includegraphics{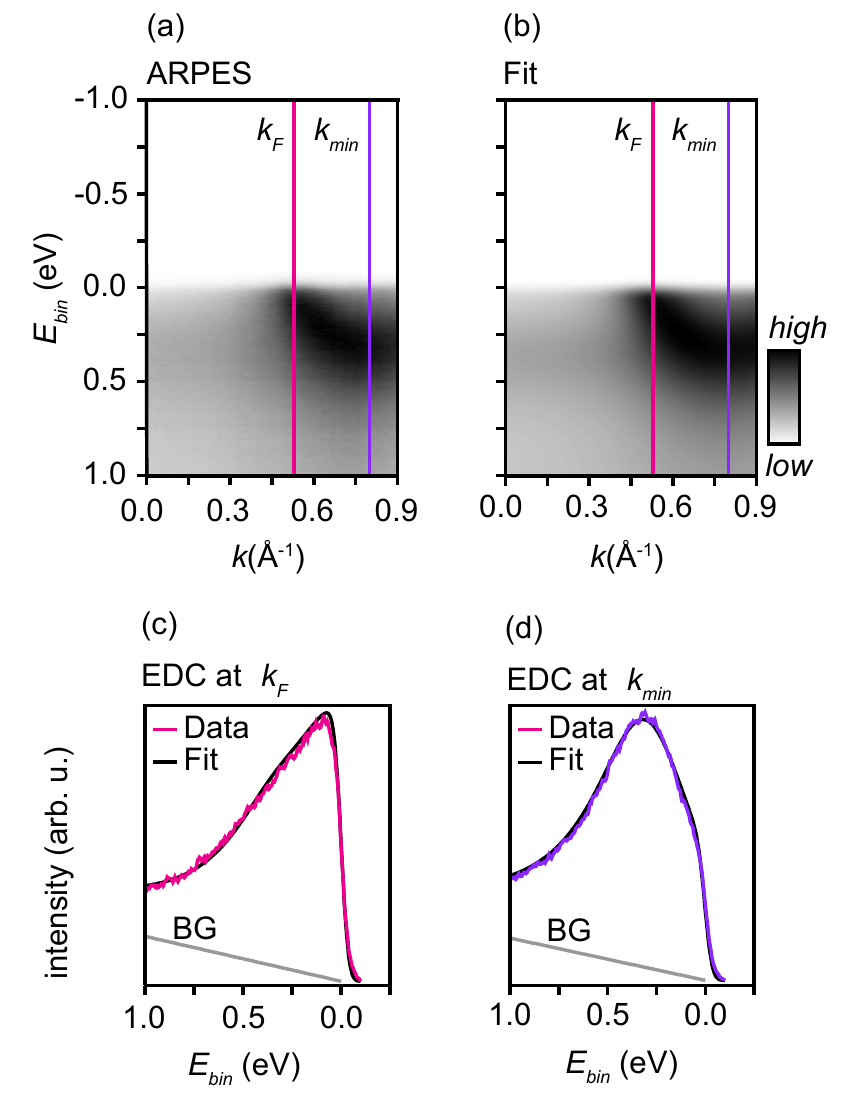}\\
\caption{(Color online) (a) Static ARPES data of the SL TaS$_2$ parabolic state with the band minimum located at $k_{min}$ and the Fermi level crossing at $k_F$. (b) Modelled intensity over the measured region of $(E,k)$-space shown in (a). 
(c)-(d) Example EDCs of the measured data and intensity fit taken along the dashed vertical lines shown in (a)-(b) at (d) $k_F$ and (e) $k_{min}$, respectively. The background intensity in the fit is shown as a light gray line marked "BG".}
  \label{fig:2}
\end{figure}

In order to determine the equilibrium dispersion parameters, this model for the photoemission intensity is fitted to high-resolution experimental data taken at ASTRID2 at a photon energy of 25 eV, \emph{i.e.} the same photon energy as used for the probe pulse in the TR-ARPES experiment. Fig. \ref{fig:2} shows the resulting excellent agreement between (a) measured and (b) modelled spectral function. 
Fig. \ref{fig:2}(c) and (d) further show the degree of agreement in the form of EDC-cuts at the Fermi level crossing ($k_F$) and at the band minimum ($k_{min}$). In this fit, the highest binding energy reaches a value of 350~meV. The light gray lines in these cuts represent the background (BG) function that is not included in equ. (\ref{eqn:4}). 


\subsection{(iii) Extracting electronic temperature and band shifts}

\begin{figure*}
\begin{center}
\includegraphics{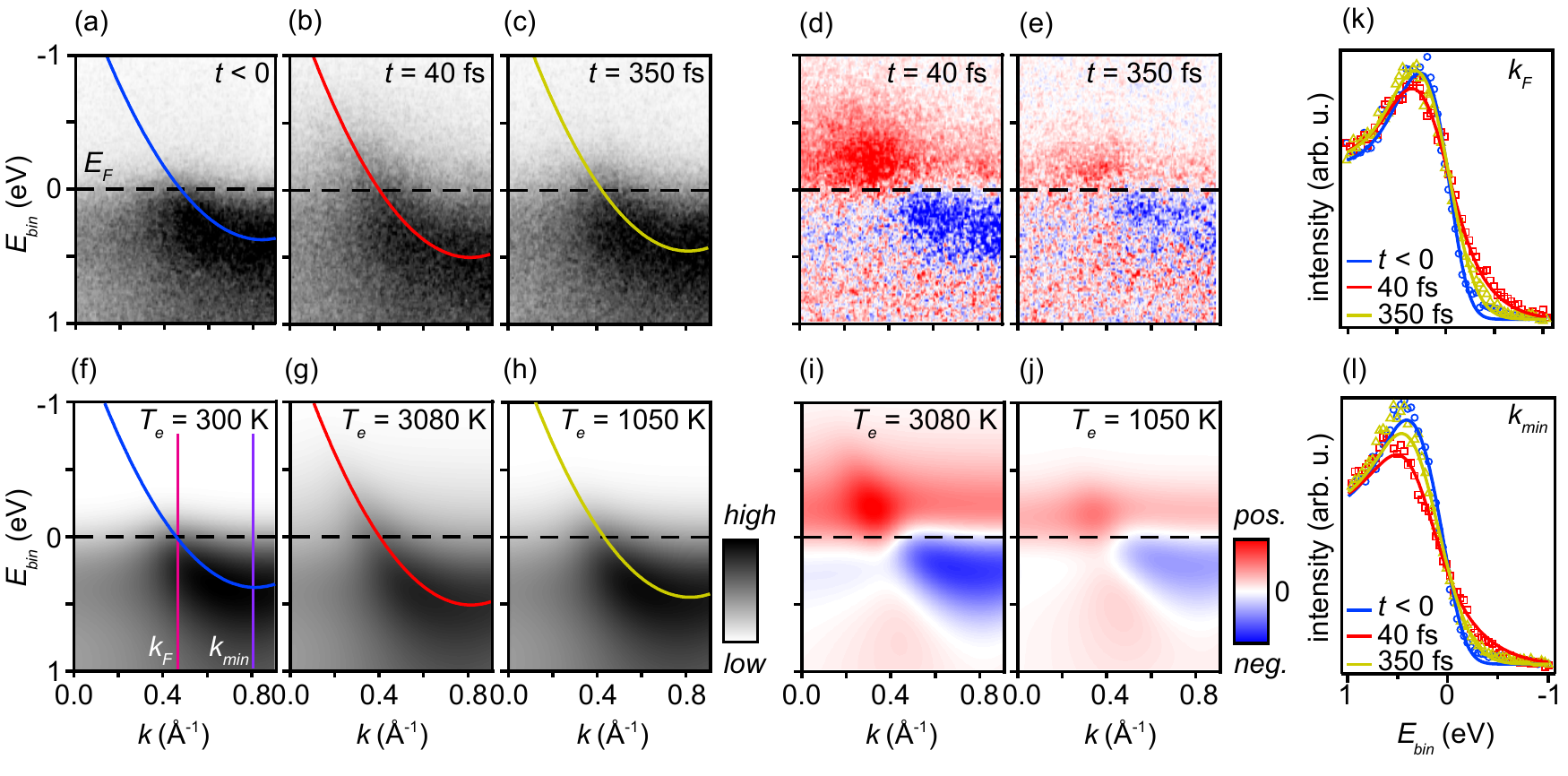}
\caption{(Color online) Time dependence of the excited-state signal in SL TaS$_2$ and spectral function simulations: (a)-(c) TR-ARPES data obtained at the given time delays for an optical excitation energy of 2.05~eV and a pump laser fluence of 7.8~mJ/cm$^2$ with the sample at a temperature of 300~K. The spectrum in (a) was taken before optical excitation. The fitted parabolic dispersions derived according to equ. (\ref{eqn:4}) are shown on top of the spectra and coloured to distinguish the different time delays. (d)-(e) Difference spectra determined by subtracting the equilibrium spectrum in (a) from the excited state spectra in (b)-(c). (f)-(j) Simulated intensity (difference) corresponding to the measured data in (a)-(e). (k),(l) Comparison of EDCs from measurements (symbols) and simulations (lines) at $k_{F}$ and $k_{min}$, respectively (see pink and purple lines in (f)).}
\label{fig:3}
\end{center}
\end{figure*}

The approach of fitting $\mathcal{I}_{conv}(E,k)$ to a model spectral function now permits the precise determination of parameters such as the electronic temperature $T_e$ and changes of the dispersion. Fig. \ref{fig:3} shows representative results for the application of this fitting method to time-resolved data sets. Fig. \ref{fig:3}(a)-(c) show the measured dispersion in equilibrium, at maximum excitation ($t=$40~fs) and at $t=$350~fs, while  Fig. \ref{fig:3}(f)-(h) show modelled spectral functions, obtained as described in section (ii) above. In order to obtain these fits, we have varied the electronic temperatures $T_e$,  the dispersion offset $c$ in equ. (\ref{eqn:4}), and the constant and linear coefficient of the linewidth $\alpha$ and $\beta$. A redistribution of background intensity following photoexcitation is also taken into account \cite{suppl}. The application of this fitting procedure results in an excellent description of the data in Fig. \ref{fig:3}(a)-(c) for all experimental parameters (delay time, fluence). This is illustrated in the comparison between measured (Fig. \ref{fig:3}(d) and (e)) and fitted (Fig. \ref{fig:3}(i) and (j)) intensity differences for time delays of 40 and 350~fs and in the direct comparison between measured and fitted EDC-cuts at $k_F$ and $k_{min}$, as shown in Fig. \ref{fig:3}(k) and (l). Fits of similar quality are obtained for data sets taken at different sample temperatures and pump laser fluence \cite{suppl}. Given the high quality of the fits obtained, we conclude that the hot electron gas is always observed to be in thermal equilibrium with a well-defined temperature, as might be expected given the time resolution of the experiment.


\begin{figure}
\begin{center}
\includegraphics{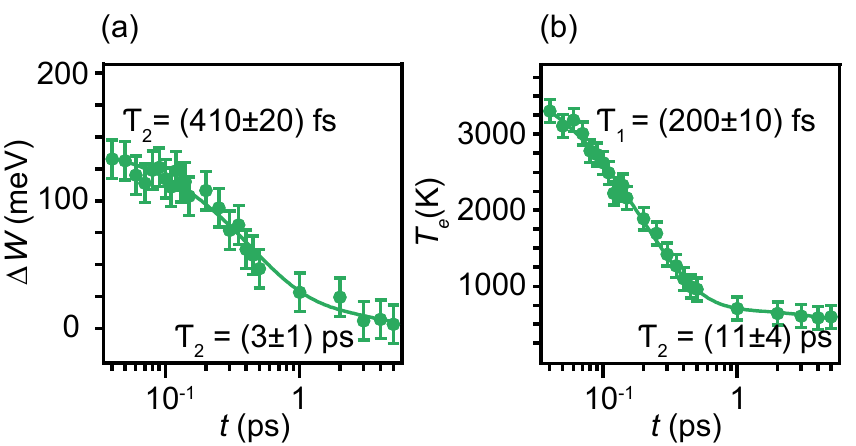}
\caption{(Color online) Extracted parameters from the data set shown in Fig. \ref{fig:3}. (a) Time dependence of the extracted band shift  $\Delta W$. The fit to a double exponential function is shown (solid line) and the relaxation times $\tau_{1}$ and $\tau_{2}$ are given. (b) Corresponding data and fit for the electron temperature $T_e$.}
\label{fig:4}
\end{center}
\end{figure}

The extracted dynamic changes of the dispersion and electronic temperature $T_e$ are given in Fig. \ref{fig:4}. In order to quantify the band shift, we introduce the quantity $\Delta W$,  defined as the difference between the band minimum energy in the excited and equilibrium state, that is the change of the parameter $c$ in equ. (\ref{eqn:4}). Fig. \ref{fig:4}(a) shows that this shift is very substantial -  more than 100~meV at peak excitation (\emph{i. e.} roughly a third of the total occupied bandwidth). Fig. \ref{fig:4}(b) shows the shift takes place at the same time delays as the peak temperature of the electron gas exceeding 3000~K. The time dependence of both $\Delta W$ and $T_e$ can be described by a double-exponential decay with a relaxation time $\tau_1$ well below 1~ps and a slower $\tau_2$. We tentatively assign  $\tau_1$ to a decay process involving the excitations of high energy optical phonons in the TaS$_2$ layer \cite{Nicki_2017} and the slower decay to a combination of acoustic phonon excitations and anharmonic decay of the optical phonons, similar to the situation seen in graphene \cite{Johannsen:2013ab}. 

This presence of a transient hot electron distribution is thus accompanied by a substantial band shift that is not \emph{a priori} expected for a metallic system. Note that the fit was constrained to a rigid band shift, without other changes and, in particular, without the boundary condition of fixed $k_F$ values. This procedure was chosen because the position of the band minimum is an experimentally well accessible quantity, as this spectral region is least affected by the broad Fermi-Dirac distribution. Moreover, the change in $k_F$ resulting from a rigid shift of the dispersion is small ($<0.05$~\AA$^{-1}$) and would be much harder to resolve than the energy shift.  Assuming merely a rigid shift of the band also appears justified because of the high fit quality at all time delays. However, minor changes in the dispersion and of the Fermi wave vector cannot be completely excluded.

\begin{figure}
\begin{center}
\includegraphics{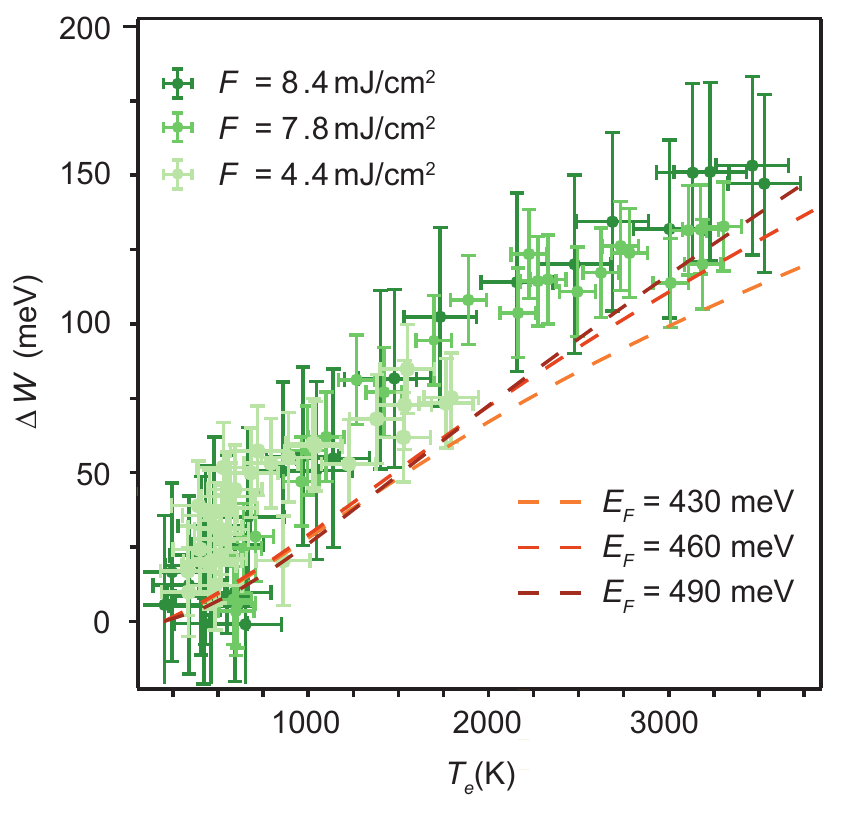}
\caption{(Color online) Temperature-induced band shift, using data for different choices of laser fluence and sample temperature. The data points show the extracted experimental band shift $\Delta W$ as a function of electronic temperature $T_e$. The curves are the calculated change in the occupied bandwidth as a function of $T_e$ for three different values of the system's Fermi energy (chemical potential at $T_e=0$.)}
\label{fig:5}
\end{center}
\end{figure}

A strong correlation between $\Delta W$ and $T_e$ emerges when we combine all data points obtained at different fluences and sample temperatures and plot $\Delta W$ as a function of $T_e$ in Fig. \ref{fig:5}.  The cause of this correlation is explored in the next sub-sections. However, the relation $\Delta W(T_e)$ is clearly not strictly linear, explaining the fact that the relaxation times $\tau_1$ and $\tau_2$ from the fit of the data in Fig. \ref{fig:4}(a) and (b) are not necessarily the same. 

\subsection{(iv) Calculated chemical potential shifts}

While the observed correlation between $T_e$ and $\Delta W$ does not imply causality, it is tempting to seek a simple mechanism that can explain the band shift as caused by the high electronic temperature without invoking, for example, substrate effects. In the present section we explore how the band shift can be caused by a temperature-induced shift of the chemical potential, which is required to conserve the total charge in the system. In the following sub-section we also address the possibility of additional shifts caused by the temperature-dependence of the electronic screening. 
\par
Small changes of the chemical potential compared to the zero temperature Fermi energy are always expected in a metallic system, but since $k_B T$ in typical experimental conditions is much smaller than the Fermi energy the shifts are also small. This is clearly not the case here. Indeed, the width of the Fermi-Dirac function at 3000~K exceeds the occupied bandwidth of  SL TaS$_2$, and temperature-induced shift of the chemical potential could thus be considerable. Predicting the size or even the direction of the shift is not trivial because it involves the details of the entire occupied and unoccupied electronic band structure. 
\par
\begin{figure} [t!]
\begin{center}
\includegraphics{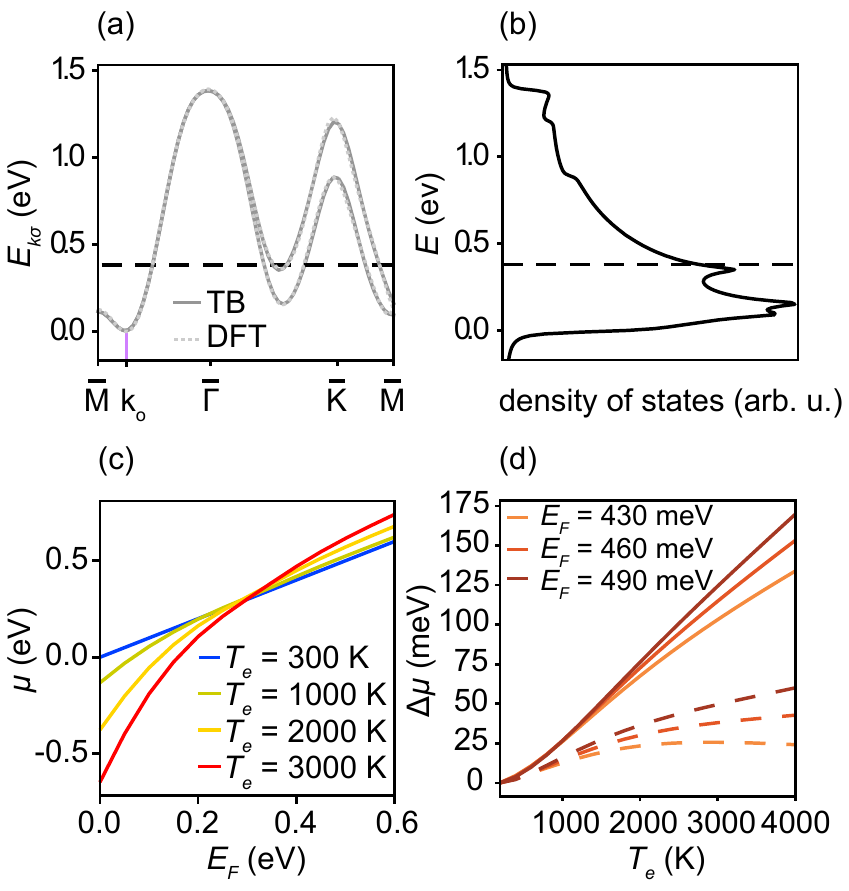}
\caption{(Color online) (a) Calculated single-particle dispersion of the topmost valence band of SL TaS$_2$. The solid line is the result from a tight-binding calculation with parameters fitted to a density functional theory calculation (dashed line). 
Note that the bands are spin-split at $\overline{K}$.  The dashed black line shows the position of the Fermi energy for a filling of the band with one electron per unit cell, as expected for the free-standing layer. (b) Resulting density of states. (c) Chemical potential versus the Fermi energy at different values of $T_e$ with $\Gamma_{\bf k\sigma}=10$meV and $\Sigma_{\bf k\sigma}=0$.   
(d)The chemical potential shift versus $T_e$ for three different values of the Fermi energy.  Solid (dashed) curves correspond to the absence (presence) of self-energy effects on the chemical potential.}
\label{fig:6}
\end{center}
\end{figure}

In order to calculate the expected shift of the chemical potential, we start from a tight-binding (TB) model for SL TMDCs \cite{Liu:2013ag} and adapt it for the case of SL TaS$_2$. The model is based on the $d$-orbitals of Ta atoms on a triangular lattice where the first, second and third nearest neighbour hopping integrals are taken into account. Accurate TB parameters are obtained by fits to the results of a density functional theory (DFT) calculation \cite{suppl}. The resulting single particle dispersion at zero temperature is shown in Fig. \ref{fig:6}(a) and the corresponding density of states in Fig. \ref{fig:6}(b). 
\par
Based on the known electronic structure, we can determine the temperature-dependent chemical potential using the following procedure. We start from the assumption of a fixed number of electrons $N$ at any temperature. This number is given by 
\begin{align}\label{eq:particles}
N=\int^{\infty}_{-\infty} \frac{\rho(E)}{e^{\beta(E-\mu)}+1} dE,
\end{align}
where  $\beta=1/k_{\rm B} T_e$ and  $\mu$ is the chemical potential. The density of states, including lifetime broadening effects, is given by $\rho(E)=  L^{-2}\sum_{\bf k\in \rm BZ} {\cal A}(E,{\bf k})$, where $L$ is the side length of the sample. The spectral function $ {\cal A}(E,{\bf k})$ is  given by
\begin{align}\label{eq:sepct}
{\cal A}(E,{\bf k}) = \frac{1}{\pi} \sum_{\sigma=\pm} \frac{\Gamma_{\bf k\sigma}}{(E-{\cal E}_{\bf k\sigma})^2 + \Gamma^2_{\bf k\sigma}}, 
\end{align}
in which  the quasiparticle dispersion follows 
\begin{align}\label{eq:renorm}
{\cal E}_{\bf k\sigma} =  E_{\bf k\sigma} +\Sigma_{\bf k\sigma}.
\end{align}
Note that $E_{\bf k\sigma}$ stands for the bare dispersion as a function of $\bf k$ and for a particular spin-split branch of $\sigma=\pm$ in the whole BZ \cite{suppl,Sanders2016}. $\Sigma_{\bf k\sigma}$ is the real part of the on-shell self-energy correction, $\Sigma_{\bf k\sigma} = {\rm Re}[\Sigma(E,{\bf k},\sigma)]_{E\to E_{\bf k\sigma}-\mu}$, and  $\Gamma_{\bf k\sigma}=-{\rm Im}[\Sigma(E,{\bf k},\sigma)]_{E\to  E_{\bf k\sigma}-\mu}$ is the corresponding imaginary part of the self energy resulting from electron-electron interactions at finite temperature. Note that the theoretical linewidth $2\Gamma_{\bf k\sigma}$ cannot be compared directly to the experimental linewidth. The experimental linewidth is expected to be larger, due to the contributions of electron-phonon and electron-defect scattering not present in the theoretical model. Moreover, the experimental linewidth at higher binding energies is affected by the presence of azimuthal disorder in the sample. 
\par
When defining  the absolute band minimum in the BZ $E_{k_\circ}$ as the zero of the energy scale (see Fig. \ref{fig:6}(a)), the temperature-dependent occupied bandwidth measured in the experiment is given by
\begin{align}
W(T_e) = \mu(T_e)- \Sigma_\circ(T_e),
\end{align}
where $\Sigma_\circ(T_e)$ is the real part of the self-energy at the band minimum position $k_\circ$.  $W(T_e) $ thus has two contributions, one from the temperature-dependent chemical potential and one from the electron-electron interaction that affects the energy of the band minimum via the self-energy $\Sigma_\circ(T_e)$ (we tacitly assume that self-energy effects do not move the band minimum away from $k_\circ$. This is confirmed by the calculations below). 
The temperature-induced change of the bandwidth when the system is heated from the equilibrium temperature $T_{eq}$ to $T_e$ then reads
\begin{align}
\Delta W(T_e)= W(T_e)-W(T_{eq}) = \Delta\mu(T_e)-\Delta \Sigma_\circ(T_e)
\label{eq:deltaW}
\end{align}
where  $\Delta\mu(T_e)= \mu(T_e) - \mu(T_{eq}) \approx \mu(T_e)-E_{\rm F}$  and $\Delta \Sigma_\circ(T_e)= \Sigma_\circ(T_e)- \Sigma_\circ(T_{eq})$. 
Note that $\Delta\mu(T_e)$ is also affected by self-energy effects (if present), since these can lead to a rigid shift of the entire band which is then compensated by a change in the chemical potential. This will be discussed in the next sub-section.
\par
We perform the energy integral in equ. (\ref{eq:particles}) analytically and then numerically integrate over the BZ.  
From this, the temperature-dependent chemical potential can be extracted, and thereby also the shift of the observed dispersion. In the initial iteration, we neglect the influence of many-body effects on the band dispersion by setting $\Sigma_{\bf k\sigma}=0$ and assume a constant electronic lifetime broadening of $\Gamma_{\bf k\sigma}$. Fig. \ref{fig:6}(c) shows the resulting shift of the chemical potential as a function of electron filling in the layer, expressed in terms of the Fermi energy (\emph{i.e.}, the chemical potential at zero temperature).  Evidently, the degree of change strongly depends on the position of the Fermi energy. For a very low filling of the band a temperature increase leads to a decrease of the chemical potential, whereas the opposite is the case for high filling. 
\par
These opposite trends at different filling levels can be understood by considering the following thermodynamical identity in a fixed system volume \cite{marder}: 
\begin{align}
\frac{\partial \mu }{\partial T_e}\Big |_{N} = - \frac{\partial N }{\partial T_e}\Big |_{\mu} \left( \frac{\partial N }{\partial \mu}\Big |_{T_e}\right)^{-1}~.
\end{align}
Then, utilizing the low-$T_e$ Sommerfeld expansion for the number of particles, $N$, we have \cite{marder}
\begin{align}\label{eq:dmudT}
\frac{\partial \mu }{\partial T_e}\Big |_{N}  \approx - \frac{\pi^2}{3} k^2_{\rm B} T_e \frac{\rho'(\mu)}{\rho(\mu)}~,
\end{align}
where $\rho'(\mu) = d\rho(\mu)/d\mu$ stands for the derivative of the density of states.  At low-$T_e$ we can approximate $\rho'(\mu)\approx \rho'(E_{\rm F})$. By looking at the density of states in Fig. \ref{fig:6}(b), we can see that $\rho'(E_{\rm F})$ has a positive (negative) sign in low (high) doping. This can roughly explain $ {\partial \mu }/{\partial T_e}<0$ ($ {\partial \mu }/{\partial T_e}>0$) for low (high) filling as depicted in Fig. \ref{fig:6}(c).

It is not entirely clear what choice of $E_F$ is most appropriate for a comparison with the experiment. The band minimum  at $k_{min}$ determined from the static high-resolution experiment is 350~meV. However, due to the azimuthal disorder this is not equal to the Fermi energy but roughly to the average highest binding energy in the $\overline{\Gamma}-\overline{M}$ and $\overline{\Gamma}-\overline{K}$ directions. The actual Fermi energy corresponds to the highest binding energy along $\overline{\Gamma}-\overline{M}$ (marked $k_\circ$ in Fig. \ref{fig:6}(a)) which is $\approx$80~meV higher than the average highest binding energy. A choice of $E_F=430$~meV should thus be a good estimate of the Fermi energy in the experimental data. 

We plot the temperature-induced shift in the chemical potential as a function of $T_e$ for three different values of the Fermi energy in Fig.~\ref{fig:6}(d). Due to the extreme temperatures reached and the small bandwidth, the effect of a temperature-induced chemical potential shift is essential in order to explain the observed band shift $\Delta W$. However, due to the potentially strongly temperature-dependent screening of the Coulomb interaction, a considerable contribution from the self-energy correction is also expected. This is discussed in the following section.  

%
\subsection{(v) Calculated effect of static screening}
It is conceivable that many-body effects also contribute to the observed band shift in addition to the effect stemming from the chemical potential. It is well-known, for instance, that the electronic self-energy has a significant effect on the observed dispersion of electronic states in ARPES, even in the case of simple metals \cite{Jensen:1985aa,Northrup:1987aa}. In the present experiment the electronic temperature is changed over such a wide range that it is relevant to ask if the temperature-induced change in electronic screening could contribute significantly to the observed changes in the electronic structure. 
\par
We investigate this by calculating the temperature-dependent static screened exchange self-energy using the single-band TB model of SL TaS$_2$. For this we first evaluate the density-density susceptibility which is given by \cite{Giuliani:2005}
\begin{align}
&\rchi({\bf q}, T_e)  =  \frac{1}{L^2}
\sum_{{\bf k}\in{\rm BZ}} \sum_{\sigma} \frac{f(E_{\bf k\sigma},T_e)-f(E_{{\bf k}+{\bf q }\sigma},T_e)}{E_{\bf k\sigma}-E_{{\bf k}+{\bf q}  \sigma}}.
\end{align}
%
The effect of temperature-dependent screening is treated through the static screened exchange which is given by  \cite{Giuliani:2005}
\begin{equation}\label{eq:sigma_k}
\Sigma_{\bf k\sigma}(T_e)  = -  \frac{1}{L^2}\sum_{\bf q\in {\rm BZ}}   \frac{v_{{\bf q}-{\bf k}}}{1-v_{{\bf q}-{\bf k}} \rchi({\bf q}-{\bf k},T_e)} f( E_{{\bf q}\sigma},T_e),  
\end{equation}
 where the bare Coulomb interaction in 2D reads $v_{\bf q} =  {2\pi e^2}/({\epsilon_{\rm eff}|\bf q|})$ with $\epsilon_{\rm eff} \sim (1+\epsilon_{\rm sub})/2$. Note that $\epsilon_{\rm sub}\sim 22$ is the dielectric constant of the graphene/SiC substrate \cite{Walter_prb_2011}. 
 
\begin{figure} [t!]
\begin{center}
\includegraphics{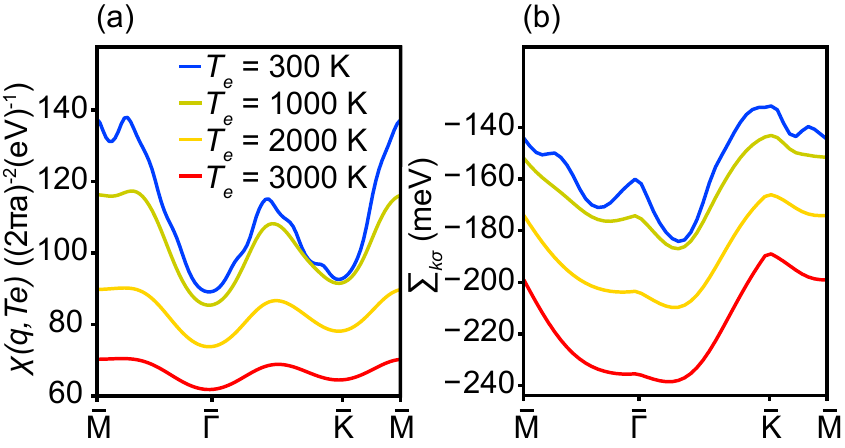}
\caption{(Color online) (a) Calculated temperature-dependent electronic susceptibility along high-symmetry lines of the BZ for four different values of $T_e$. 
(b) Real part of the electronic self-energy along high-symmetry lines of the BZ ($T_e$ as in (a)). Note that the difference between up and down spins is negligible for this set of parameters (\emph{i.e.} $\Sigma_{\bf k \uparrow} \sim \Sigma_{\bf k \downarrow}$). Note that for this figure we set $E_{\rm F}=430$meV.}
\label{fig:chi_self}
\end{center}
\end{figure}

The $q$-dependent susceptibility of SL TaS$_2$ is shown in Fig. \ref{fig:chi_self}(a) 
for four different temperatures along high-symmetry directions in the BZ and for  $E_{\rm F}=430$~meV, as observed in the experiment.
 As seen in Fig. \ref{fig:chi_self}(a), the absolute magnitude of the static susceptibility (considering that the static susceptibility is always a negative value, \emph{i.e.} $|\rchi({\bf q},T_e)|=-\rchi({\bf q},T_e)$), and therefore the screening, decreases by increasing $T_e$. This is a universal trend at low temperature and low momentum. At the BZ center and at low electronic temperature ($T_e \ll T_{\rm F}$), we can approximate $\rchi(0,T_e) \approx - \rho(\mu)$ \cite{Giuliani:2005} and by using equ.~(\ref{eq:dmudT}) we have  
 \begin{align}
\frac{\partial |\rchi(0,T_e)|}{\partial T_e}\Big|_{N} \approx \frac{\partial \mu}{\partial T_e}\Big|_{N} \rho'(\mu) <0~,
 \end{align}
 with $T_{\rm F}$ being the Fermi temperature.
 In general this negative slope is not fulfilled at finite $\bf q$ or very high temperature. Under our experimental conditions, the electronic temperature is very high, leading to a considerable broadening of the Fermi-Dirac distribution function which in fact becomes similar to the entire bandwidth (\emph{i.e.} $k_{\rm B}T_e \sim E_{\rm F}$). 
 This strong broadening reduces the probability of virtual transitions, $E_{\bf k\sigma}\to E_{{\bf k}+{\bf q}\sigma}$, for all values of $\bf q$ in the BZ. Because of this semi-Pauli-blocking effect at very high temperature, the number of virtual electron-hole excitations is diminished, leading to a weaker screening effect. A similar reduction of screening is predicted to emerge in the parabolic-band two dimensional electron gas (2DEG) \cite{Maldague_1978,Hwang_prb_2009,Sarma_rmp_2011}, \emph{i.e.} $\rchi_{\rm 2DEG}({\bf q},T_e\gg T_{\rm F})\approx T_{\rm F}/T_e $ \cite{Hwang_prb_2009}.
On the other hand, in the massless Dirac fermion (MDF) model of graphene the screening effect is predicted to have an increasing trend with temperature as $\rchi_{\rm MDF}({\bf q},T_e\gg T_{\rm F})\approx \log(4)  T_e/T_{\rm F}$\cite{Hwang_prb_2009}. This can be due to the inter-band transitions in graphene which do not exist in our single-band metallic system.
\par
Having a weaker screening of the Coulomb interaction at high electronic temperature implies a stronger many-body effect. As depicted in Fig.~\ref{fig:chi_self}(b), the explicit self-energy calculation indicates a very strong temperature dependence, with changes of $\approx$100~meV over the experimental temperature range. However, this self-energy is mostly comprised of a  rigid shift of the whole band which is compensated by an opposite shift in the chemical potential in order to conserve the particle number. When these many-body effects are included in equ. (\ref{eq:sepct}) and the chemical potential is calculated in a second iteration, this new estimate is thus strongly modified. This explains the large difference between the chemical potential shifts with and without self-energy corrections in Fig. \ref{fig:6}(d). Experimentally, neither the change in the chemical potential nor in the self-energy is directly observable but only their combination in $\Delta W$, see equ. (\ref{eq:deltaW}). Fig.~\ref{fig:5} thus compares the full theoretical result to the experimental data, showing an almost quantitative agreement between experiment and calculation. 
A comparison of the theoretical results in Fig. \ref{fig:5} and  \ref{fig:6}(d) shows that including static screening to first order leads to a bandwidth change of merely  $\sim15$~meV at $T_e\sim 4000$~K. Note that while the experimentally observed highest binding energy of the band is not equal to the occupied bandwidth, the \emph{change} in this binding energy is the same as the theoretically calculated $\Delta W$.

\section{Conclusions}
We have demonstrated strong electronic heating and changes in the occupied bandwidth  upon optical pumping of a 2D metal, SL TaS$_2$. The data could be quantitatively analyzed using a 2D fitting scheme of the entire resolution-broadened spectral function. The  experimentally observed band shifts are explained by considering the temperature-dependent many-body screening effect and the chemical potential shift required to conserve charge neutrality in the presence of a hot electron population. 
\par
The possibility of very large band shifts in pumped metallic systems could potentially be used to create a number of unconventional states of matter. We emphasize that neither the direction nor the magnitude of the shift is trivial but both result from the material's band structure in a wide range around the Fermi energy. Indeed, much larger shifts still could be expected for SL TaS$_2$ with a different band filling. Starting from an appropriate band structure, it could thus be possible to use transient temperature-induced shifts in order, for example, to push a Van Hove singularity in the density of states close to the chemical potential, possibly creating electronic instabilities at high temperatures. 

\section{Acknowledgements}
We thank Phil Rice and Alistair Cox for technical support during the Artemis beamtime. We gratefully acknowledge funding from VILLUM FONDEN through the Young Investigator Program (Grant. No. 15375) and the Centre of Excellence for Dirac Materials (Grant. No. 11744), the Danish Council for Independent Research, Natural Sciences under the Sapere Aude program (Grant Nos. DFF-4002-00029, DFF-6108- 00409 and DFF-4090-00125), the Aarhus University Research Foundation, The Leverhulme Trust and The Royal Society. Access to the Artemis Facility was funded by STFC.
%
%
%

\end{document}